# Unveiling the effect of adding B$_4$C at the W-on-Si interface

Adele Valpreda [*,a], Hendrik W. Lokhorst [a], Jacobus M. Sturm [a], Andrey E. Yakshin [a], Marcelo Ackermann [a]

[*] Corresponding author. *E-mail address*: a.valpreda@utwente.nl

*Full postal address*: University of Twente, Faculty of Science and Technology, Carré 2.013, P.O. Box 217, 7500 AE Enschede, The Netherlands

[a] Industrial Focus Group XUV Optics, MESA+ Institute for Nanotechnology, University of Twente, Enschede 7522NB, the Netherlands

# Abstract

In this study, we investigate the W-on-Si interface and the effect of adding a B$_4$C interlayer at such interface, using low-energy ion scattering (LEIS) spectroscopy, X-ray reflectivity, X-ray diffraction, and transmission electron microscopy with energy dispersive X-ray spectroscopy. We extract the effective width of the interface in three different structures having: no-B$_4$C, 0.24 nm, and 1.2 nm of B$_4$C deposited in between the W and the Si films. The analysis reveals that the W distribution does not get significantly sharper when B$_4$C atoms are deposited at the W-on-Si interface, showing that B$_4$C does not act as a physical barrier against the diffusion of atoms during the deposition of these structures.

W/Si thin-film structures are used in various applications, including X-ray optics. While many studies reported that adding a sub-nm thick B$_4$C film at the W/Si interfaces is beneficial for the overall reflectivity of the structures, the physical mechanisms involved were not yet fully understood. In this context, being able to characterize the width of the interfaces with sub-nm resolution is key. In this study, we show how the analysis of the sub-surface signal of the LEIS spectra enables the characterization of the interface W-on-(B$_4$C-)Si, highlighting the importance of systematically extracting the values of the effective interface width for the analysis and understanding of thin-film growth.

# 1. Introduction

Structures of W/B$_4$C and W/Si thin films are used in multilayer systems for the reflection of soft X-rays [1-3]. These multilayer systems are stacks of deposited thin films that act as a Bragg reflector [3]. To fulfill Bragg's law in the soft X-ray wavelength range (λ = 2.36–0.834 nm), the films must be very thin (< 3 nm), and this creates engineering challenges. When using W/$B_4C$, the roughness of the interfaces is one of the main problems. The study [3] showed that this can be addressed by applying low-energy ion beam polishing during deposition. The authors reported a 43% peak reflectance at a wavelength of λ = 0.834 nm achieved with a W/$B_4C$ multilayer where a 50 eV ion beam polishing was applied [3]. When using W/Si, the formation of W silicides at the interface between the thin films is one of the main problems. These silicides lower the optical contrast between the W and the Si, and this reduces the soft X-ray reflectance [1, 2, 4]. Therefore, several techniques have been applied in order to reduce the W-Si interaction [1, 2, 5]. Specifically, following the findings on the effects of $B_4C$ as a thermal diffusion barrier for Mo/Si multilayers [6, 7], the study [2] showed that adding a few monolayers of sputtered $B_4C$ at the W/Si interfaces leads to a multilayer with higher soft X-ray reflectivity performances, compared to the W/Si and W/$B_4C$ reference structures. The authors reported a 45% peak reflectance at a wavelength of λ = 0.834 nm achieved with a Si/$B_4C$/W multilayer, where a sub-nm thick $B_4C$ film was deposited at the W-on-Si interface [2].

While the literature is rich in studies reporting the increase in reflectivity obtained by the addition of a thin $B_4C$ layer in the stack [2, 8-13], the physical mechanisms involved are not yet fully understood. In [2], the authors performed grazing incidence x-ray reflectivity and x-ray fluorescence measurements on the multilayer stack. The results indicate that adding $B_4C$ at the W-on-Si interface leads to a higher concentration of W in the absorber layer (the W film). Furthermore, the authors performed XPS measurements, which showed that when adding $B_4C$ at the W-on-Si interface, the W-Si bonds are partially replaced by either W-W bonds or W-$B_4C$ bonds [2].

As a follow-up, in this study, we systematically measure the width of the W-on-Si interface for three structures having no-$B_4C$, 2.4 Å, and 1.2 nm of $B_4C$ by depth-resolved low energy ion scattering. Furthermore, to confirm the ability of our metrology technique to resolve the small changes at the interface of interest, we characterize the same samples also by X-ray reflectivity (XRR), x-ray diffraction (XRD), and transmission electron microscopy (TEM).

In Low Energy Ion scattering (LEIS) measurements, a beam of ions of a noble gas is directed toward the sample, and the final energy of the backscattered ions is measured. While the technique is known for its surface selectivity of the very topmost atomic layers of the sample, depth-resolved information of the composition of the first few 10 nm of the samples can also be present in LEIS spectra. In recent years, we have developed an approach to resolve thin film interfaces buried in the first few nanometers of the sample by extracting the depth-resolved information from LEIS spectra, [14-16]. This is interesting because it avoids the use of sputter steps to probe the interface, and it extends the use of LEIS spectra to the measurements of those materials that are subjected to matrix effects, such as $B_4C$ [17].

In [15], we investigated the Si-on-W interface using low energy ion scattering (LEIS), X-ray reflectivity (XRR), and transmission electron microscopy (TEM). By extracting the Si-on-W effective interface width from the three different techniques, we demonstrated that, in the case of Si-on-W, the effective interface width measured by LEIS and XRR agrees with the values extracted from TEM analysis within a 0.1 nm error margin. Considering the extensive experimental effort required by TEM, these results exemplified the value of LEIS and XRR as analysis techniques for resolving thin film interfaces. However, we still needed to test the applicability of the developed approach to resolve a complex interface that is nevertheless interesting for X-ray and EUV optics.

In this study, we show how the analysis of the sub-surface signal of the LEIS spectra enables the characterization of the interface W-on-($B_4$C-)Si. Furthermore, our analysis reveals that the W distribution at the W-on-Si interface does not get sharper when $B_4$C atoms are deposited at the interface, showing that $B_4$C does not act as a physical barrier against the diffusion of atoms during the deposition, confirming the findings in the study [2], which suggested that adding $B_4$C leads to the formation of compounds that are chemically more stable and have better contrast for X-ray reflectivity.

## 2. Experimental methods

For this study, we deposited three samples that we analyzed using LEIS, XRR, XRD, and TEM. All samples contain a 15 nm thick amorphous Si underlayer, a 2.2 nm thick W film, and a 8 nm thick Si capping layer. The 8 nm thick Si capping layer was deposited after the LEIS analysis of the W films and has the purpose of protecting the W film from oxidation while the sample is exposed to air. The schematic in Fig. 1 displays the deposition and analysis process. The X-ray measurements were performed in air on the capped structures. The TEM measurements were performed on lamellae extracted from the sample using a focused ion-beam (FIB) method.

All films were deposited by magnetron sputtering on Si wafer pieces. The deposition chamber used for this study is an in-house designed ultrahigh vacuum (UHV) system with a base pressure < $1 \times 10^{-9}$ mbar. All thin films were deposited at room temperature. Argon was used as sputter gas for all depositions. The working pressure was around $1 \times 10^{-3}$ mbar.

The thickness of the W and Si film was monitored during deposition by quartz crystal microbalances (QCM) that were calibrated with XRR measurements. This allows us to measure the as-deposited thickness, taking into account the possible fluctuations of the deposition rate from the nominal value. The thickness readings measured by the QCM deviate less than 5 % from the XRR-measured thickness.

For the first sample, the W film was deposited directly on the Si underlayer. For the second and third samples, a 2.4 Å layer of $B_4C$ and a 1.2 nm layer of $B_4C$ were deposited before the deposition of the W film, as indicated in the schematic of Fig. 1. The reported thickness of the $B_4C$ layers is a nominal as-deposited value, which was calculated from the XRR measurement of a thicker $B_4C$ calibration sample. The 2.4 Å (0.24 nm) layer of $B_4C$ in the second sample virtually represents a monolayer of $B_4C$ atoms deposited at the interface W-on-Si. The 1.2 nm layer of $B_4C$ in the third sample represents a higher amount of $B_4C$ deposited at the interface W-on-Si.

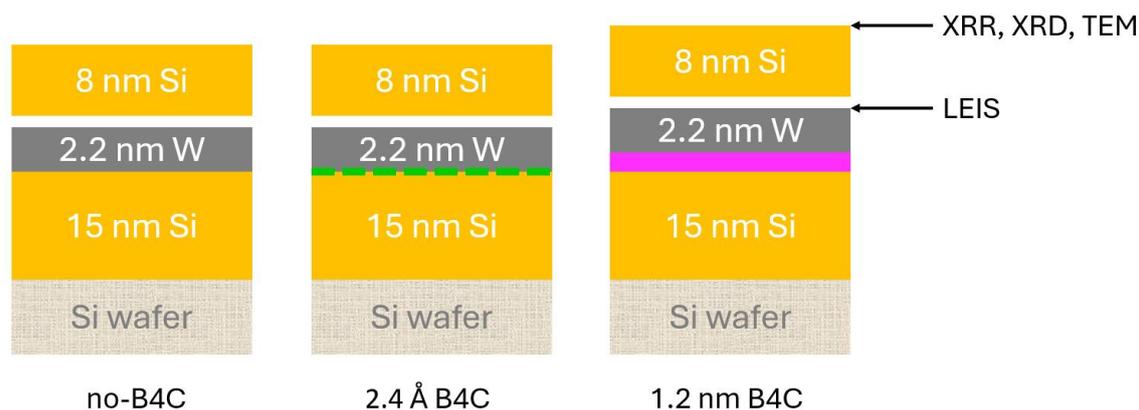

*Fig. 1. Schematic of the deposition and analysis process. All films were deposited by magnetron sputtering. In each sample, a 15 ± 1 nm thick Si film was deposited on the Si wafer. This deposited Si film represents the underlayer for the other films in the structures. In the sample named no-B4C, a 2.2 ± 0.1 nm thick W film was deposited directly on the Si underlayer. In the sample named 2.4 Å B4C, the equivalent to a monolayer of B4C atoms was deposited on the Si underlayer before the 2.2 ± 0.1 nm thick W film was deposited. In the sample named 1.2 nm B4C, a layer ca. 1.2 nm thick of B4C atoms was deposited on the Si underlayer before the 2.2 ± 0.1 nm thick W film was deposited. These samples were transferred in vacuum to the LEIS chamber and LEIS spectra were recorded. After the LEIS measurements, each sample was transferred back to the deposition chamber, where an 8 ± 0.5 nm thick Si film was deposited as a capping layer. After that, the samples were exposed to air, where the x-ray reflectivity, x-ray diffraction, and transmission electron microscopy measurements were performed.*

The LEIS measurements were performed in-vacuum in a Qtac 100 spectrometer from IONTOF after less than 3 min of in-vacuum transfer from the deposition chamber. In our setup, the primary gun - generating the ion beam for the measurement - is perpendicular to the sample's surface. A double toroidal electrostatic analyzer accepts ions (not neutrals) that are backscattered at an angle of 145° with respect to the sample's surface normal and directs them toward a position-sensitive detector. During a LEIS measurement, a primary beam of He ions scans over a 1×1mm$^2$ area. Measurements were performed with a primary beam energy of 3 keV. In all the measurements, the beam current was in the range of 2-4 nA, and the acquisition time was under 4 min with an ion dose of around 3 × 10$^{14}$ ions/cm$^2$. The measured LEIS spectra plot the yield of ions backscattered at an angle of 145° as a function of the ion's final energy. For the simulation of the LEIS spectra, TRBS simulations of backscattered particles were multiplied by an experimentally determined ion fraction. For the exact description of the simulation procedure, we refer to our previous publications [14, 16].

The XRR measurements were performed on a Malvern Panalytical Empyrean diffractometer using a hybrid monochromator that provides a Cu Kα1 beam (1.54056 Å). The beam was about 0.07 mm wide, producing a line focus. The measured XRR curves plot the reflected intensity of an X-ray beam as a function of the incidence angle with respect to the sample surface. The detector angle is specular to the incident angle. For the simulation of the XRR curves, the IMD software was used [18]. For the reconstruction of the XRR curves, the AMASS software by Malvern Panalytical was used with the Free-Form model developed by Yakunin et al. (2014) [19].

The XRD measurements were performed on a Malvern Panalytical X'Pert diffractometer system. Before each measurement, the sample was rotated 20 degrees around its axis (normal to the sample's surface) to reduce the detected diffraction from the Si wafer. Note that the crystal lattice of the wafer is aligned with the sample's edges. During a measurement, the angle between the incoming beam and the sample's surface (ω) was fixed at 1 °. This represents a grazing incidence configuration, which maximizes the illuminated area and minimizes penetration into the substrate. The position of the detector varied during the measurements. Specifically, the angle between the incoming beam and the detector (2θ) went from 5 ° to 135 °. The resulting XRD graph plots the intensity of the diffracted beam as a function of the 2θ angle. A schematic of the angles with respect to the sample in an XRD measurement is reported in [20].

The transmission electron microscopy (TEM) cross-sectional images were taken using a Thermo-Scientific probe-corrected Spectra 300 TEM system at an acceleration voltage of 300 kV. Lamellae for TEM were extracted from the samples using a focused ion beam (FIB) method. A carbon film was deposited on the surface of the sample before the FIB cutting to protect the sample from damage. The electron beam was aligned along the Si(110) zone axis of the <100> silicon wafer. TEM bright field (BF) images, scanning-TEM High angle annular dark field (HAADF) images, and EDX analysis were taken. For the analysis of the EDX data, the Velox microscope operation software on Spectra 300 was used.

## 3. Results and discussion

### 3.1. Low Energy Ion Scattering (LEIS) Spectroscopy

LEIS spectra recorded with a 3 keV He ion beam are presented in Fig. 2. The three spectra show a single surface peak at an energy of around 2650 eV, corresponding to the final energy of He ions scattered by W atoms at θ = 145°. These spectra result from structures whose surface is homogeneously covered by W atoms. At an energy of around 2500 eV and below, the spectra show a broader and lower intensity signal. This signal, typically described as the "tail" of the surface peak, comes from the He ions scattered by subsurface W. In this specific case of a 2.2 nm thick W film, this part of the spectra (shaped like an error function) directly corresponds to the signal coming from the interfacial W. At an energy of around 1500 eV and below, the rising of the yield corresponds to the depth signal from He ions scattered by Si atoms.

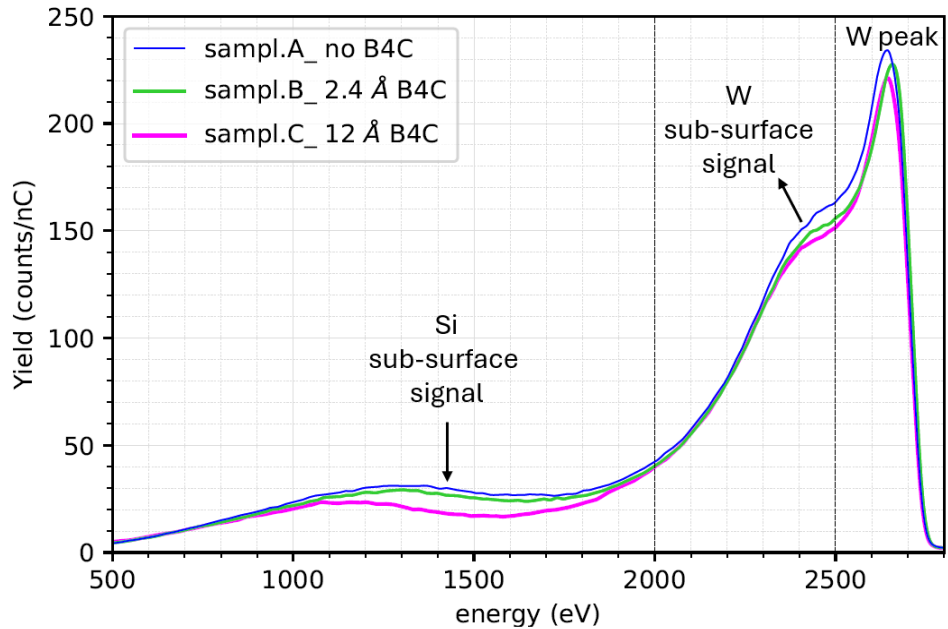

*Fig. 2. LEIS spectra of the structures having no-B4C, ca. 2.4 Å of B4C, and ca. 1.2 nm of B4C deposited at the W-on-Si interface. The position of the W surface peak, the W sub-surface signal, and the Si sub-surface signal are marked on the plot. The two vertical dashed lines mark the area of the spectra that is shaped like an error function, which corresponds to the signal coming from the interfacial.*

Fig. 2 shows that the effect of adding $B_4C$ at the interface is twofold: 1) the Si signal gets pushed to lower energies, and 2) the yield of the W sub-surface signal decreases. The fact (1) that the silicon signal gets pushed to lower energy is expected; by adding an extra layer between the W and the Si, the He ions have to travel a longer path inside the sample before being backscattered by Si. The longer the path, the more energy losses there are. This highlights the sensitivity of LEIS spectra to the depth composition. To interpret the decrease in the yield of the W sub-surface signal (effect 2), we proceed by simulating the LEIS spectra of several structures having different widths of the W-on-Si interface. For the exact description of the simulation procedure, we refer to our previous publication [16].

In Fig. 3, we present a qualitative comparison of the simulated LEIS spectra (fig. 3a) and the experimental LEIS spectra (fig. 3b) recorded with a 3 keV He ion beam. Note that these simulations can be directly compared with the experimental spectra because they are the product of a simulated spectrum of backscattered particles and an experimentally determined ion fraction for backscattering from a W surface.

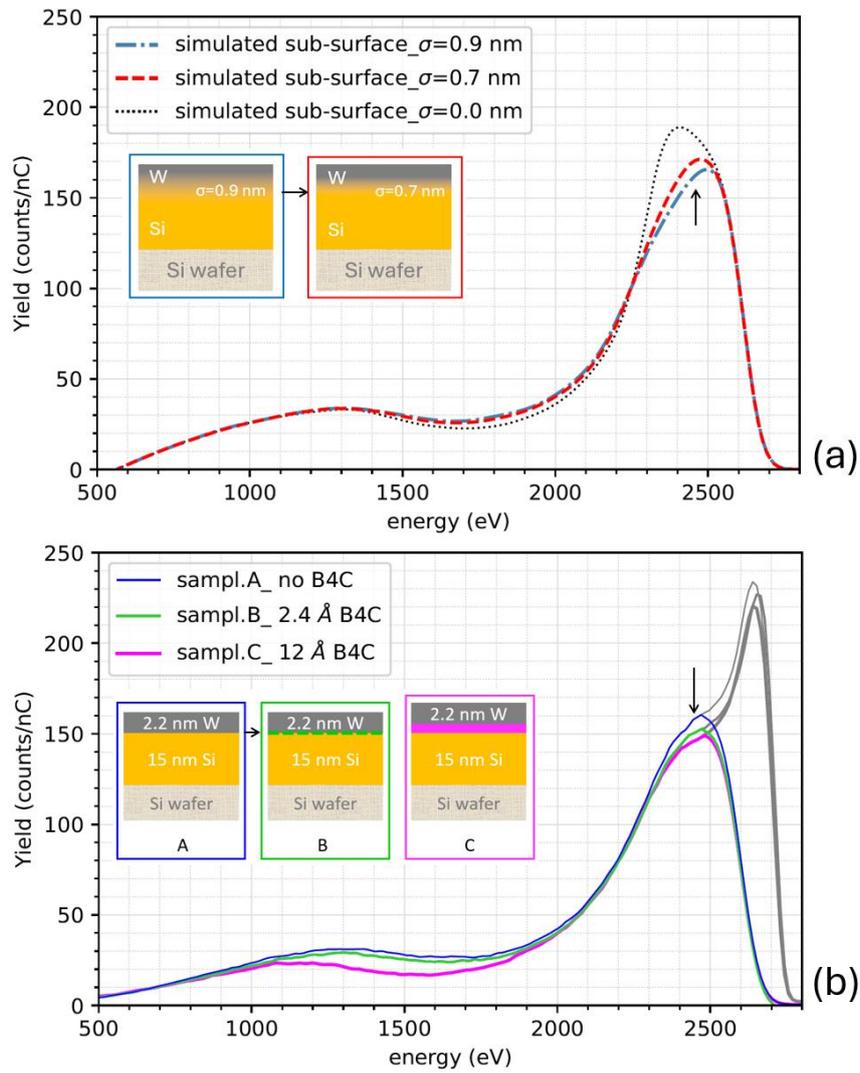

*Fig. 3. The simulated spectra in (a) are compared to the experimentally measured LEIS spectra in (b). In (a), three simulated spectra are displayed to show that the sharper the W-on-Si interface, the higher the yield of the W sub-surface signal. In (b), the sub-surface signals of the experimentally measured LEIS spectra show that adding B4C at the W-on-Si interface reduces the yield of the W sub-surface signal. The two black arrows on the plots are added to point at the described trend. The grey lines show the experimentally measured LEIS spectra from which the sub-surface signals (colorful lines) are extracted by subtraction of the surface peak.*

The simulations in Fig. 3a indicate that, for a sample having 2.2 nm of W on Si, when the W-on-Si interface sharpens, from a width of σ=0.9 nm to a width of σ=0.7 nm, the yield of the W sub-surface signal increases. This is due to the higher concentration of W in the top film. The experimental spectra in Fig. 3b show that when adding a monolayer (nominal thickness 2.4 Å) of $B_4C$ atoms at the W-on-Si interface, the yield of the W sub-surface signal decreases and that when adding a $B_4C$ layer of nominal thickness 1.2 nm, the yield of the W sub-surface signal decreases even more. This indicates that the atoms of $B_4C$ deposited at the interface W-on-Si do not act like a physical barrier to the intermixing of the W and Si during deposition.

To extract values of the effective interface width from the experimental spectra, we proceed by fitting the signal from interfacial W. For the 3 keV LEIS spectra of the no-$B_4C$ sample, an effective interface width σ=0.9 ± 0.1 nm resulted in the best fit of the experimental data [16]. We repeated the same fitting process also for the 3 keV LEIS spectra of the other two structures, having a B$_4$C film of nominal thickness of 2.4 Å and 1.2 nm, respectively, deposited at the interface. The results are shown in Fig. 4.

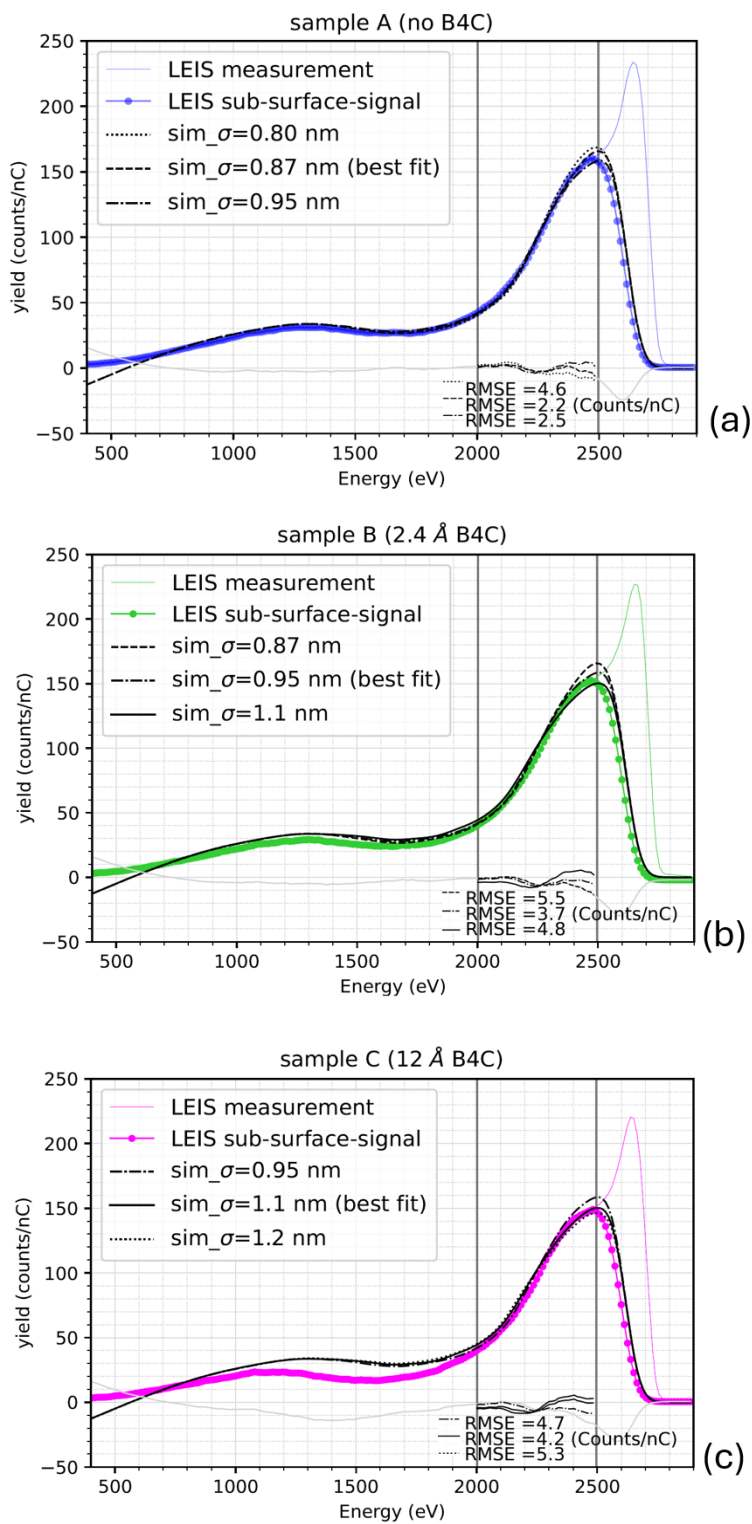

Fig. 4. LEIS spectra and corresponding simulations for the three structures investigated in this study. For each LEIS measurement, the sub-surface signal is highlighted as a thick dotted line, and it is compared to three simulated sub-surface signals. The vertical lines mark the area of the spectrum used to calculate the root mean square error (RMSE) of the fit. The simulation resulting in the minimum RMSE corresponds to the best fit to the experimental spectrum. In (a), (b), and (c), the simulated structure that corresponds to the best fit of the experimental spectrum has an effective interface width of σ=0.87 nm, σ=0.95 nm, and σ=1.1 nm, respectively.

The result in Fig. 4 shows that the W distribution at the W-on-Si interface gets broader by less than 0.1 nm when ca. 2.4 Å of B$_4$C atoms are deposited at the interface, going from σ$_A$=0.9 ± 0.1 to σ$_B$=1.0 ± 0.1 nm.

It must be noted that the surface roughness at this stage is unknown and that a measurement of the surface roughness at the top of the W film is challenging, as taking the sample out of the vacuum at this stage would change the surface by oxidation and, therefore, modify the roughness. In the supplementary material, we illustrated how surface roughness can affect the measured interface width.

### 3.2. Specular X-ray reflectivity (XRR) measurements

After the deposition of the Si capping layer and air exposure, we measured the X-ray reflectivity curve of each sample. The results are shown in Fig. 5. Since the difference between the three investigated structures is a few monolayers at the W-on-Si interface (while the thickness of the W and the Si films is the same), it is possible to directly compare the XRR curves of the three structures.

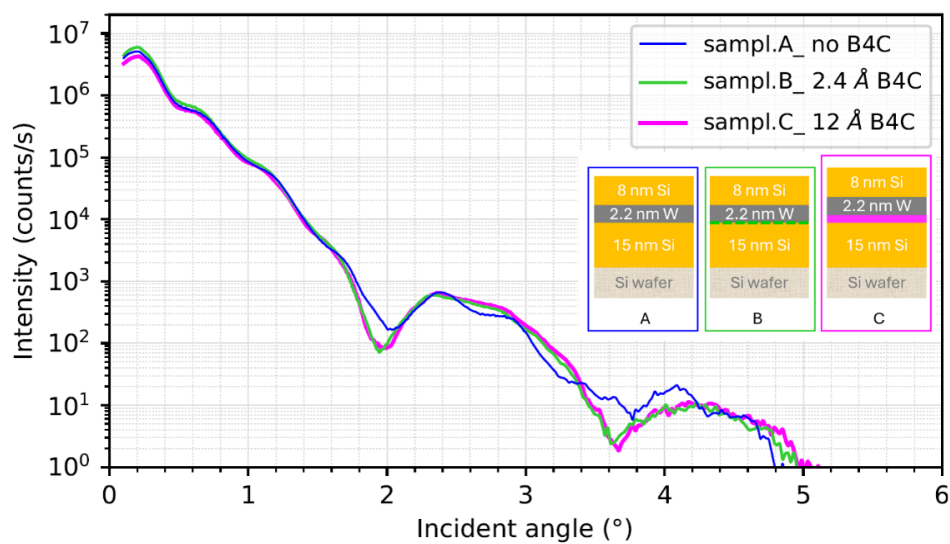

*Fig.5. Specular X-ray reflectivity (XRR) curves of the three samples investigated in this study. The intensity of the X-ray specular reflection is plotted on a logarithmic scale as a function of the incident angle. A schematic of the structures is displayed on the plot.*

The comparison of the measured XRR curves of the three samples in Fig. 5 indicates that the effect of adding B$_4$C at the W-on-Si interface is twofold: 1) an increase in the amplitude of the low-frequency oscillations (often referred to as fringes) and 2) a decrease in the amplitude of the high-frequency oscillations. Additional measurements and simulations of XRR curves - investigating the possible reasons for these changes - are presented in the supplementary material. In Fig. 6, we present the fit of the XRR curve of each sample and the corresponding simulated density profile as a function of depth in the samples.

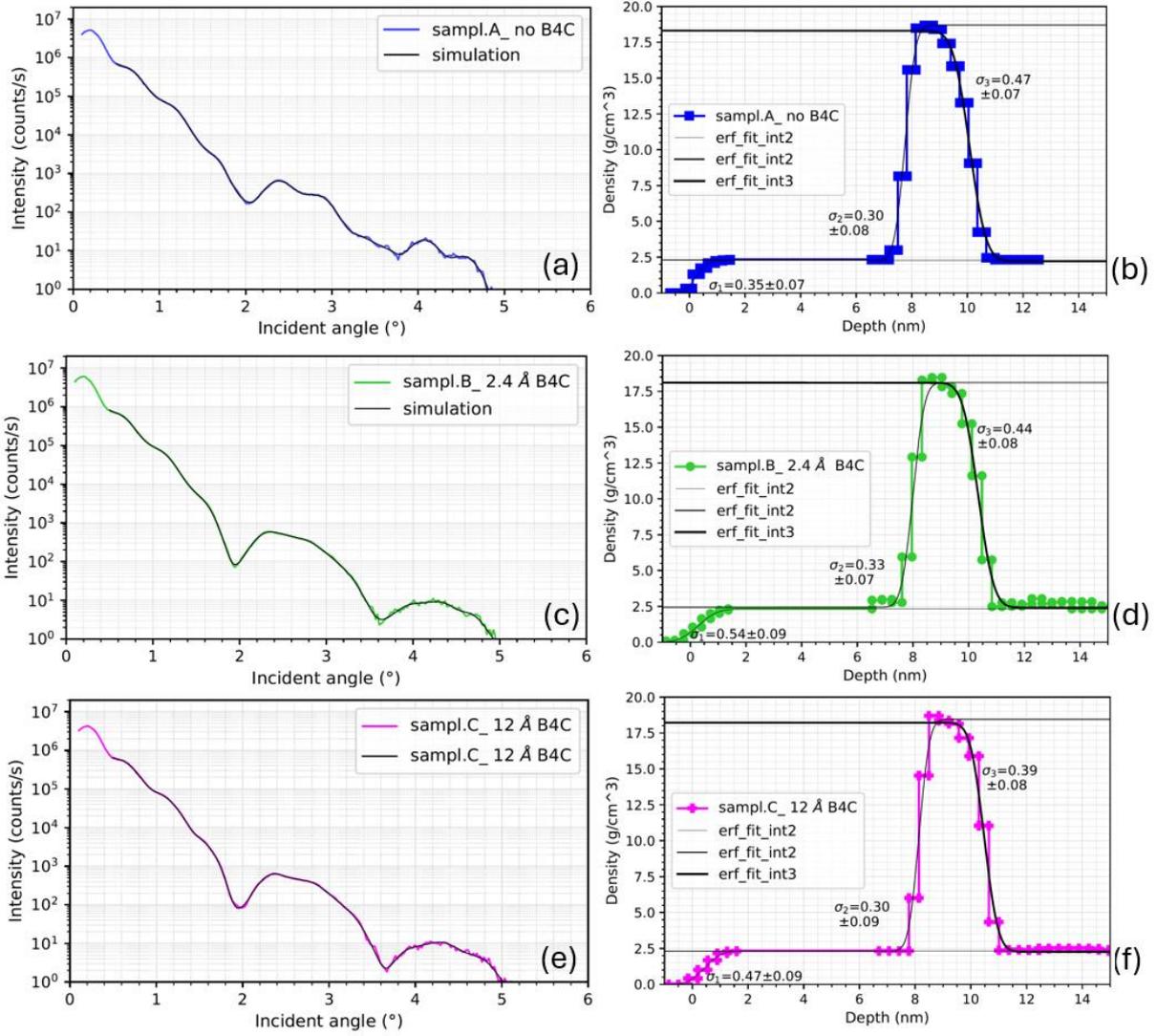

*Fig. 6. The fitted x-ray reflectivity (XRR) curves (a, c, e) and the corresponding reconstructed density profiles (b, d, f) of the three samples investigated in this study. On each density profile, the top three interfaces that are present in the structure are fitted with an error function whose parameter σ is displayed on the plot.*

The density profiles in Fig. 6 b, d, f show the top three interfaces present in the structure. By fitting the density profile at the interfaces with an error function, we can extract and compare the values of the parameter σ (in nm) describing the width of the interface [15]. The values are reported in Table 1.

*Table 1. The effective width (σ in nm) of the top three interfaces extracted from the XRR analysis. The error bar around the σ value consists of the uncertainty given by the fitting of the step-like density profiles with an error function.*

| Sample | $\sigma_1$ (Surface) | $\sigma_2$ (Si-on-W) | $\sigma_3$ (W-on-Si) |
|---|---|---|---|
| A (no-$B_4C$) | 0.35 ± 0.07 nm | 0.30 ± 0.08 nm | 0.47 ± 0.07 nm |
| B (2.4 Å $B_4C$) | 0.54 ± 0.09 nm | 0.33 ± 0.07 nm | 0.44 ± 0.08 nm |
| C (1.2 nm $B_4C$) | 0.47 ± 0.09 nm | 0.30 ± 0.09 nm | 0.39 ± 0.08 nm |

The results in Fig. 6 and Table 1 show that the interface of interest (W-on-Si) does not get significantly sharper. This is in agreement with the W concentration profiles that were obtained in [2] from the reconstruction of the GI-XRR and X-ray fluorescence curves, confirming that the effect of adding a few monolayers of $B_4C$ at the W-on-Si interface is not a physical barrier against diffusion of W atoms during deposition. Rather, the effect is the formation of compounds that have better contrast for X-ray reflectivity. This was observed in the XPS measurements performed in [2] which showed that the W-Si bonds are partially replaced by either W-W bonds or W-$B_4C$ bonds.

In Table 2, we compare the values of effective interface width that we extracted for the interface of interest (W-on-Si) from the reconstruction of the XRR curves and from the reconstruction of the LEIS spectra. The comparison shows that even for the sample having no-$B_4C$, the LEIS and XRR values are not in good agreement. One possible reason for this discrepancy was discussed in [16], where we investigated the influence of the straggling of the beam on the width of the LEIS signal. Due to the stochastic nature of ion scattering collisions, the thicker the W film, the wider the signal from the W at the bottom interface. Since the LEIS simulations underestimate straggling, the interface width measured with this method is likely to be an overestimate [15, 16].

Another consideration is that when the W surface has a high and uncorrelated roughness this can result in a broadening of the signal from the W-on-Si interface (as discussed in the supplementary material). It is also possible that the input parameters that we used to describe the W film (in the LEIS and XRR simulations) are not representative of a W film of such low thickness. For example, the density of thin films is usually less than in bulk materials. The comparison becomes even more challenging considering that an additional Si film was deposited on the W film before air exposure and XRR measurement.

Finally, while the values of the interface width ($\sigma$ in nm) extracted from the XRR curves and from the LEIS spectra are not in agreement, it is still important to investigate the effect of adding a few monolayers of $B_4C$ at the W-on-Si interface by systematically measuring the interface width in several structures having increasing amount of $B_4C$. Specifically, measuring the effective interface width by a complementary technique - such as LEIS - helps disentangle the different aspects that contribute to x-ray reflectivity. The LEIS analysis showed that the W distribution at the W-on-Si interface does not get sharper when $B_4C$ atoms are deposited at the interface, confirming that $B_4C$ does not act as a physical barrier against the diffusion of atoms during deposition [2].

Table 2. The effective width of the W-on-Si interface as measured by LEIS and XRR. For the LEIS buried interface profile method, the 0.1 nm error bar around the σ value is described in [15]. For the XRR analysis, the 0.1 nm error bar around the σ value consists of the uncertainty given by the fitting of the step-like density profiles with an error function.

| Method | W-on-Si | W-on-2.4 Å $B_4C$-on-Si | W-on-1.2 nm $B_4C$-on-Si |
|---|---|---|---|
| LEIS | $\sigma_A$=0.9 ± 0.1 nm | $\sigma_B$=1.0 ± 0.1 nm | $\sigma_C$=1.1 ± 0.1 nm |
| XRR | $\sigma_A$=0.5 ± 0.1 nm | $\sigma_B$=0.4 ± 0.1 nm | $\sigma_C$=0.4 ± 0.1 nm |

## 3.3. X-ray diffraction (XRD) measurements

To further investigate the effect of adding $B_4C$ at the W-on-Si interface, we proceed by checking the microstructure of the W films by X-ray diffraction. The results are shown in Fig. 7.

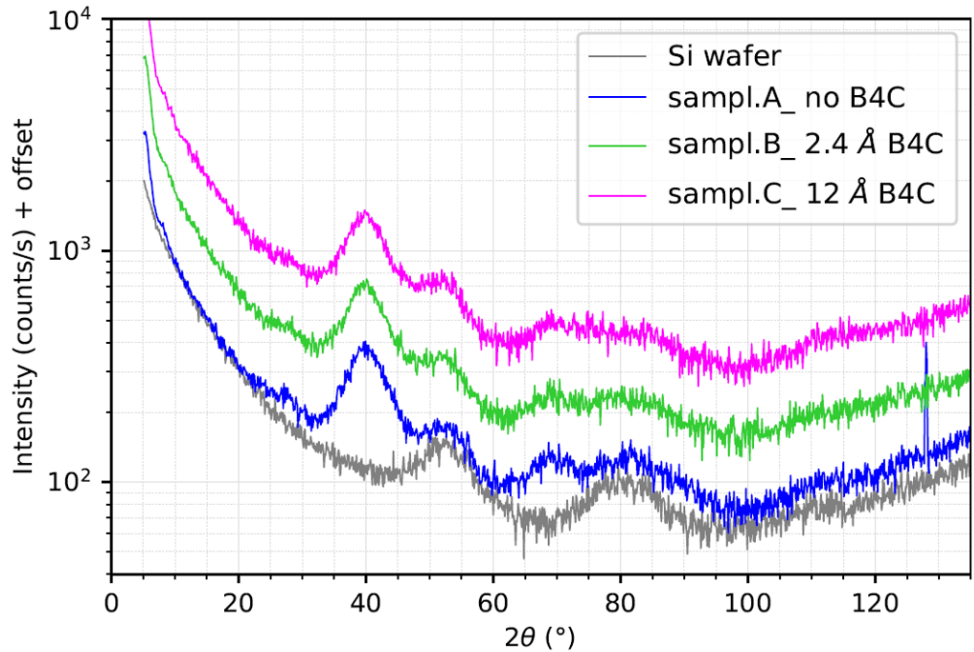

*Fig. 7. X-ray diffraction (XRD) curves of the three investigated structures. The XRD curve of a wafer was also measured with the same settings as the investigated samples and is plotted (in grey) for comparison. The angle between the incoming beam and the sample was kept constant at ω=1 °, and the angle between the incoming beam and the detector (named 2θ) varied in the range of 5 to 135 °. The intensity of the diffracted X-rays is plotted as a function of the angle 2θ. By rotating the samples 20 ° around their axis (φ), we successfully avoided the detection of the main diffraction peaks from the crystalline silicon wafer. The XRD curve of a wafer (in grey) shows the remaining signal from the Si wafer.*

The results in Fig. 7 show that the W film has a quasi-amorphous structure in all three samples investigated in this study [21-23]. Therefore, in this case of W films of thickness 2.2 nm, we can exclude a change in the microstructure of the W film as the effect of adding $B_4C$ at the W-on-Si interface.

## 3.4. Transmission electron microscopy (TEM) and Energy Dispersive X-ray (EDX) Spectrometry

After cutting a lamella of the two samples having no-$B_4C$ and 2.4 Å of $B_4C$ with the Focused Ion Beam (FIB) process, the cross-section of the lamellae was imaged by transmission electron microscopy.

Fig. 8 shows the EDX atomic fraction of each element for samples A and B compared, having no-$B_4C$ and 2.4 Å of $B_4C$, respectively. From Fig. 8, we notice a discrepancy in the measured thickness of the W film: when measured by transmission through the cross-section, the thickness of the W film is 3.8 nm higher than the thickness of the W film measured by X-ray reflectivity in the same sample (2.2 nm), indicating that the W atoms intermixed with the Si atoms during the lamella preparation. Note the XRR and TEM measurements were performed on the exact same samples and that the same discrepancy between TEM and XRR measured thickness of the W film holds for both samples having no-$B_4C$ and 2.4 Å of $B_4C$. In our previous study [15], two structures - containing respectively 4 and 20 nm of W - were characterized by XRR and TEM. In that case, the XRR and TEM measured thicknesses of the W film were in agreement with each other. A future study could investigate whether W films thinner than the crystallization threshold (c.a. 3 nm) are more likely to get intermixed with the Si film during the preparation of the lamellae for the TEM due to the amorphous and quasi-amorphous microstructure of the film. Regardless, since the lamellae for the TEM analysis of the two samples were prepared in the exact same way, a qualitative comparison between the EDX spectroscopy results of the two samples is still valid (and useful for the purpose of this study, which aims to investigate the effect of adding a few $B_4C$ atoms at the W-on-Si interface). The entirety of the TEM analysis of the two samples is included in the supplementary material for completeness.

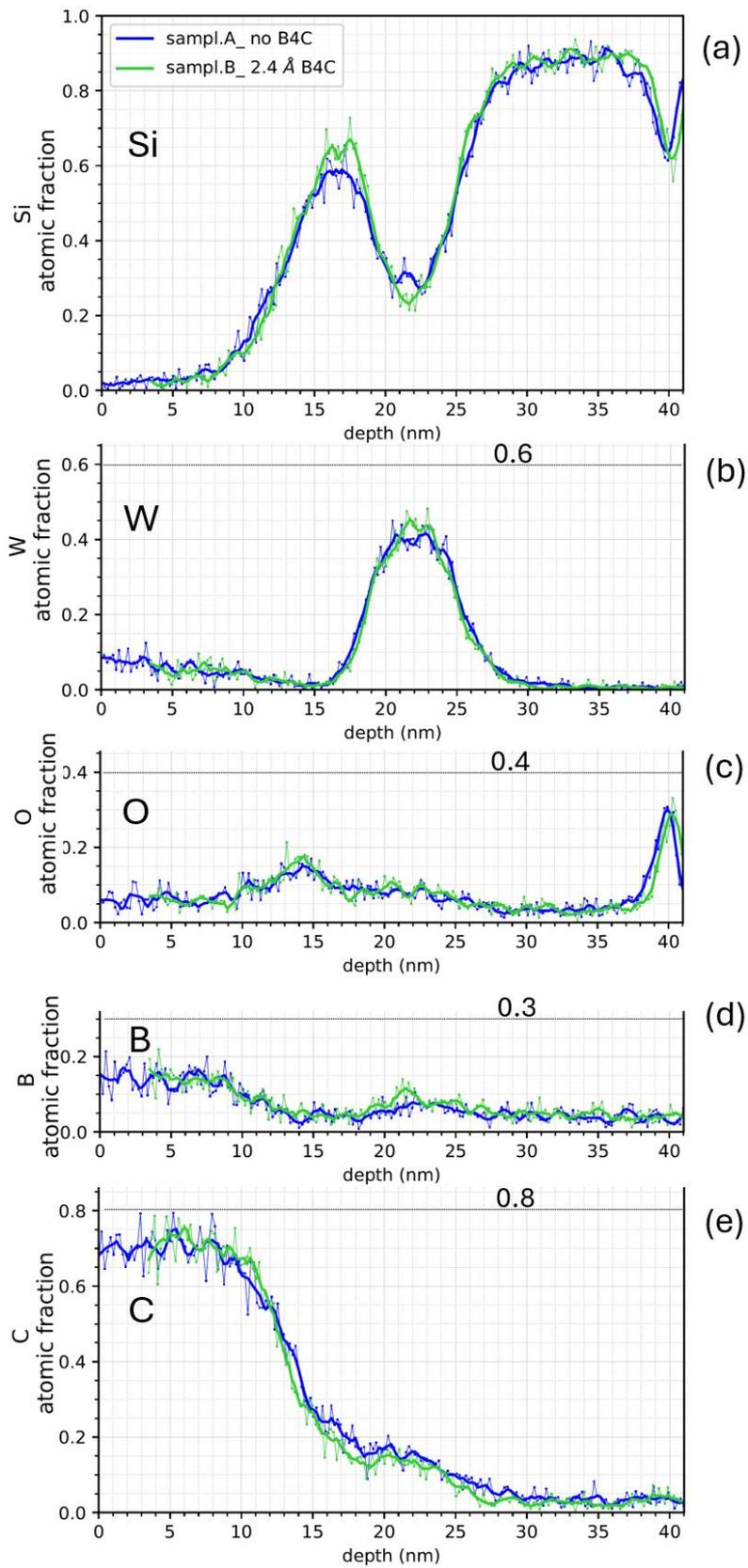

*Fig. 8. Comparison of the Energy Dispersive X-ray (EDX) spectra of the two samples having no-B4C (in blue) and 2.4 Å of B4C (in green). Each subplot describes the elemental distribution of an element in the depth of the two samples. The element name is indicated on each subplot. For the elements W, O, B, and C, a dotted horizontal line is used to highlight the value of the maximum atomic fraction displayed on the subplot.*

The results in Fig. 8a-e show that adding a few B$_4$C atoms at the interface changes the W-Si distribution in these lamellae. In Fig. 8a, the Si atomic fraction in the W film decreases, going from 30 to 25 %, while the W atomic fraction increases (Fig. 8b), going from 40 to 44%. As a consequence, the Si fraction in the top Si film increases, going from 57 to 65%.

The increase in the concentration of W in the W film due to the deposition of B$_4$C atoms at the W-on-Si interface was also observed in W/Si multilayers [2]. Note that we didn't detect any significant difference in the composition of the outermost W atomic layer from the LEIS spectra prior to the deposition of the top silicon film. A possible explanation is that the stress induced (or released) in the W film by the addition of the B$_4$C atoms changes the intermixing of Si atoms with W at the Si-on-W interface. This explains why the Si fraction increases in the Si film deposited on top of the W film. To confirm this hypothesis, LEIS measurements could be performed to probe the next interface, Si-on-W, when B$_4$C is deposited below the W film.

In Fig. 8d, we compare the measured boron atomic distribution in the two samples. Considering the "boron" atomic distribution in the no-B$_4$C sample as the background noise, a small boron peak is visible in the W film of the 2.4 Å B$_4$C sample. This indicates that the boron atoms may diffuse into the W film. The diffusion of B atoms into the transition metal film was also observed in previous studies [7, 24]. Nevertheless, more measurements are needed to verify the reproducibility of the boron signal in the samples investigated in this study.

## 4. Conclusions

By characterizing three structures with different amounts of B$_4$C at the W-on-Si interface, we investigated the effect of depositing a few monolayers of B4C at the W-on-Si interface.

The low energy ion scattering (LEIS) analysis of these structures revealed that the W-on-Si interface does not get sharper when B$_4$C atoms are deposited at the interface. By fitting the W signal in the LEIS spectra, we observe that the effective interface width goes from $\sigma_A$=0.9 ± 0.1 to $\sigma_B$=1.0 ± 0.1 nm when a monolayer of B$_4$C atoms is deposited at the W-on-Si interface and up to $\sigma_C$=1.1 ± 0.1 nm when 1.2 nm of B$_4$C are deposited at the W-on-Si interface.

By fitting the XRR curves of these structures, we reconstructed the density profile as a function of depth. The comparison of the density profiles of the three investigated structures confirmed that adding a few B$_4$C monolayers at the W-on-Si interface does not lead to a significantly sharper W-on-Si interface, going from $\sigma_{A\_XRR}$=0.5 ± 0.1 to $\sigma_{B\_XRR}$=0.4 ± 0.1 nm when a monolayer (or more) of B$_4$C atoms is deposited at the W-on-Si interface.

Possible reasons for the discrepancy between the effective interface width values extracted by LEIS and XRR were discussed. Note that the comparison was made difficult by several factors, including the fact that the W film can't be easily measured in air due to oxidation. These results highlight the importance of measuring the effective interface width by several complementary techniques. In this study, extracting the effective interface width for the three structures from the LEIS analysis showed that the W distribution at the W-on-Si interface does not get sharper when $B_4C$ atoms are deposited at the interface, confirming that $B_4C$ does not act as a physical barrier against diffusion of atoms during deposition.

The XRD analysis of the samples revealed that the W films in these structures (of thickness 2.2 nm) are quasi-amorphous and that the microstructure does not change when $B_4C$ atoms are deposited at the interface.

The transmission electron microscopy EDX analysis of lamellae from the two samples having no-$B_4C$ and 2.4 Å of $B_4C$ revealed that the W-Si distribution changes not only in the W film but also in the Si capping film that we deposited on top of the W film to prevent the oxidation of the W film during the air exposure. To further investigate whether it is the next interface to sharpen (for example, due to stress induced or released in the W film), a follow-up study could LEIS measure the "next" Si-on-W interface for these structures, having different amounts of $B_4C$ deposited at the W-on-Si interface.

For the samples investigated in this study, we noticed a discrepancy between the XRR and TEM measurements of the W film thickness. Since this discrepancy was not observed in our previous study on thicker W films (> 4 nm), we suspect that the intermixing of the W and the Si occurs during the lamella preparation when the W film is either amorphous or quasi-amorphous. To investigate this, a future study should repeat the preparation of the lamellae of the samples, for example, by using different settings. Regardless, the LEIS analyses of the structure before air exposure are extremely valuable for thin-film structures since they do not require the preparation of a cross-section of the sample.

# CRediT authorship contribution statement

Adele Valpreda: Writing – review & editing, Writing – original draft, Visualization, Methodology, Investigation, Formal analysis, Data curation, Conceptualization.

Hendrik W. Lokhorst: Methodology, Investigation, Formal analysis, Data curation.

Jacobus M. Sturm: Writing – review & editing, Data curation, Validation.

Andrey E. Yakshin: Writing – review & editing, Validation, Project administration, Data curation, Conceptualization.

Marcelo Ackermann: Writing – review & editing, Validation, Supervision, Project administration, Formal analysis, Funding acquisition.

# Declaration of competing interest

The authors declare that they have no known competing financial interests or personal relationships that could have appeared to influence the work reported in this paper.

# Data availability

Data will be made available on request.

# Acknowledgments

This work was carried out in the frame of the Industrial Partnership Program "X-tools," Project No. 741.018.301, funded by the Netherlands Organization for Scientific Research, ASML, Carl Zeiss SMT, and Malvern Panalytical. Support for the preparation of the FIB lamellae and TEM images from Melissa Goodwin and Martina Tsvetanova at the MESA + Institute for Nanotechnology of the University of Twente is gratefully acknowledged.

# Supplementary material

## On the effect of uncorrelated surface roughness on the effective interface width measured from the LEIS spectra

If the roughness at the top of the W film is uncorrelated to the roughness of the W-on-Si interface, there are points on the probed area where the W film appears thinner than the average measured value, and vice versa, as illustrated in the schematic in fig. S1.

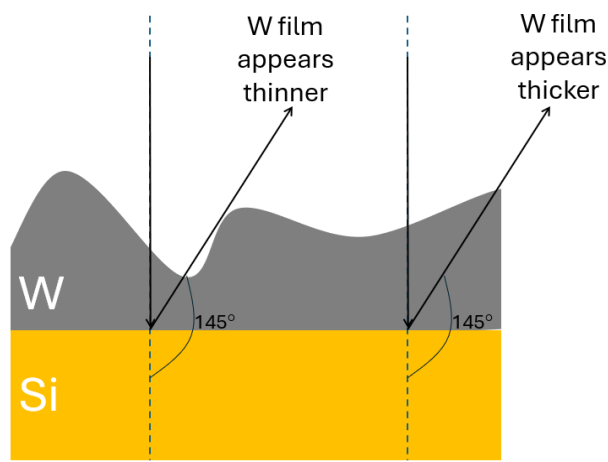

*Fig. S1. Schematic of surface roughness uncorrelated to the interface roughness. In this schematic, the interface roughness is absent, while the pronounced surface roughness affects the local thickness of the W film.*

Regardless of the lateral length scale of the surface roughness, uncorrelated roughness will always influence the path length from the surface to the interface along the incoming path. Depending on the lateral length scale of the surface roughness, the projectiles that are backscattered by W atoms at the interface may perceive an additional variation of the local thickness of the W film along the outgoing path. Variations in the path length affect the energy loss of the projectiles (the stopping) and, therefore, the final energy, resulting in a broadening of the signal from the interfacial W.

To estimate the extent of this effect, one must know the height and the lateral scale of the surface roughness. While this information is not available for the 2.2 nm W film deposited in this study, we can use the TEM images of the cross-section of a thin film as a first estimate.

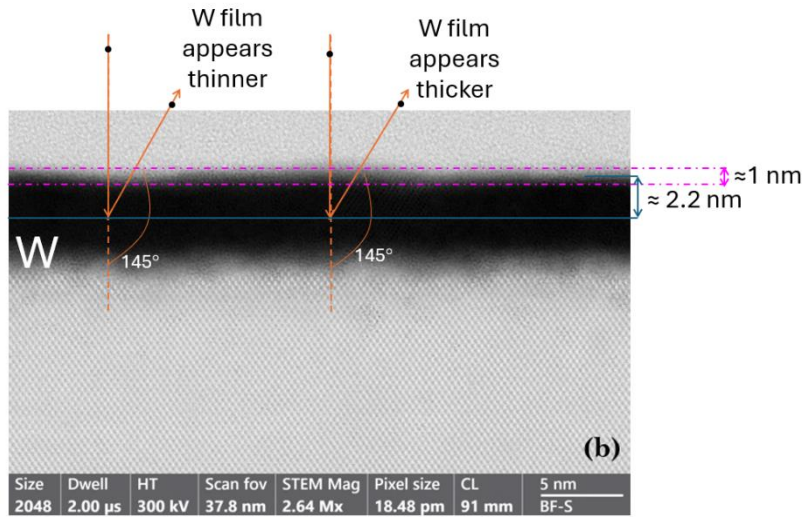

*Fig. S2. Schematic of how uncorrelated surface roughness might affect the LEIS measurements of a thin film sample. The TEM image used for the sketch was taken from [15] with permission. From this image, we recognized a grain that has a lateral size of 5 nm. The height variation on the W surface is of the order of 1 nm peak-to-valley.*

Fig. S2 shows that a surface roughness of 1 nm is a valid first estimate. Considering this value as the total width of the surface roughness (as it appears in Fig. S2), then the parameter σ of an error function describing the profile is σ=0.5 nm. This calculation indicates that surface roughness is not a negligible term when it comes to measuring the interface width from the shape of the tails in LEIS spectra.

## Additional XRR experiments and simulations

To investigate whether a simple variation in the thickness of the stack can explain the observed changes in the XRR curves, we deposited a new structure having a 16 nm Si bottom film and no $B_4C$ at the W-on-Si interface. In Fig. S3, we compare the XRR cure of the new structure to the previous data.

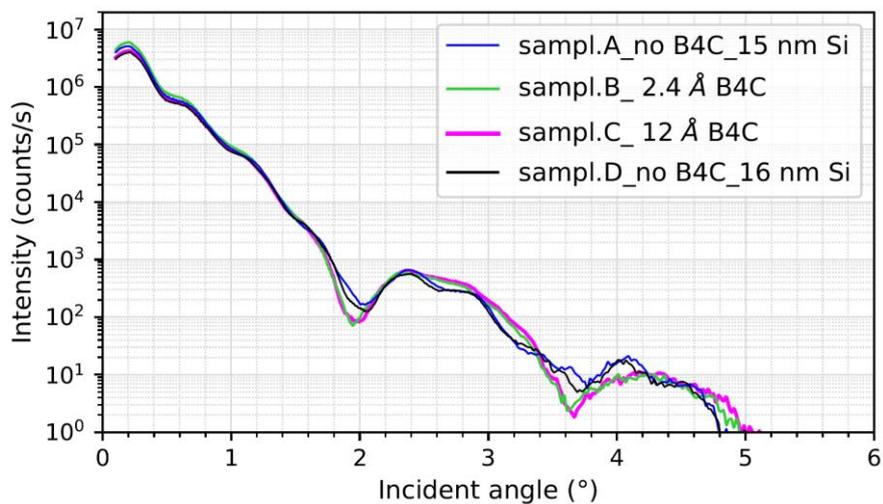

*Fig. S3. Specular X-ray reflectivity (XRR) curves of four samples. The XRR curve of a new sample (black curve), having no-$B_4C$ and 16 nm of Si as the bottom film, is compared to the three structures investigated in this study. The intensity of the X-ray specular reflection is plotted on a logarithmic scale as a function of the incident angle.*

The results in Fig. S3 show that the increase in the amplitude of the low-frequency oscillations, observed as an effect of adding a few $B_4C$ monolayers at the W-on-Si interface, can't be explained simply by a change in the interference of the x-rays caused by an increase in the thickness of the bottom film.

The simulations in Fig. S4 show that there are several factors contributing to the amplitude of the low-frequency oscillations in the measured XRR curves. The LEIS analysis of these structures revealed that the W distribution at the W-on-Si interface does not get sharper when 2.4 Å and 1.2 nm of $B_4C$ are deposited at the interface. This excludes scenario (a) of Fig. S4. The XRR curve of the new sample, having no-$B_4C$ and 16 nm of the bottom Si film, revealed that a change in the interference pattern caused by a thicker bottom film does not lead to the same increase in the amplitude of the low-frequency oscillations as the addition of a few $B_4C$ monolayers at the interface. This excludes the scenario (c) of fig. S4.

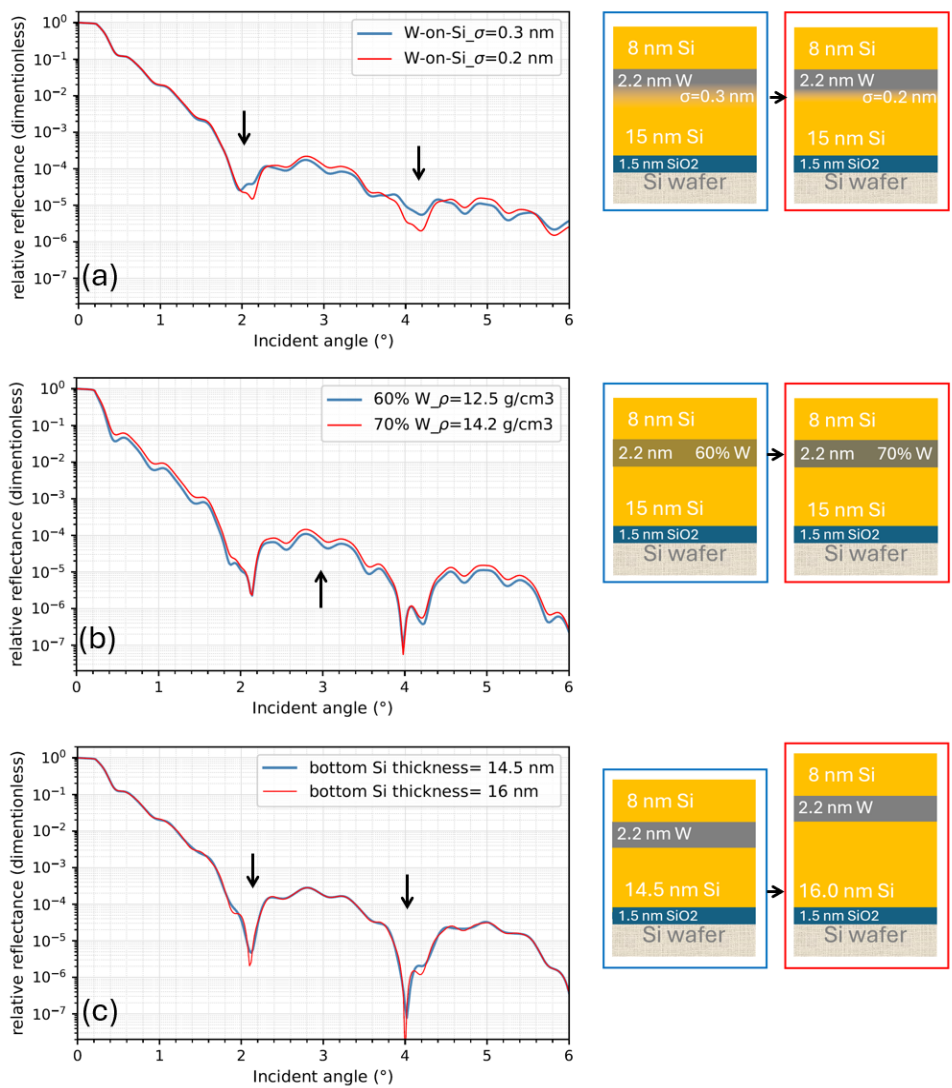

*Fig.S4. Simulated specular X-ray reflectivity (XRR) curves and the corresponding structures. In (a), the width of the W-on-Si interface is varied from σ=0.3 nm to σ=0.2 nm. In (b), the density of the W-Si film is varied from ρ=12.5 g/cm3 to ρ=14.2 g/cm3. In (c), the thickness of the bottom Si film is varied from 14.5 nm to 16.0 nm.*

# Transmission electron microscopy images and EDX analysis

Fig. S5 and S6 display the TEM analysis of the two samples having no-$B_4C$ and 2.4 Å of $B_4C$ deposited at the W-on-Si interface.

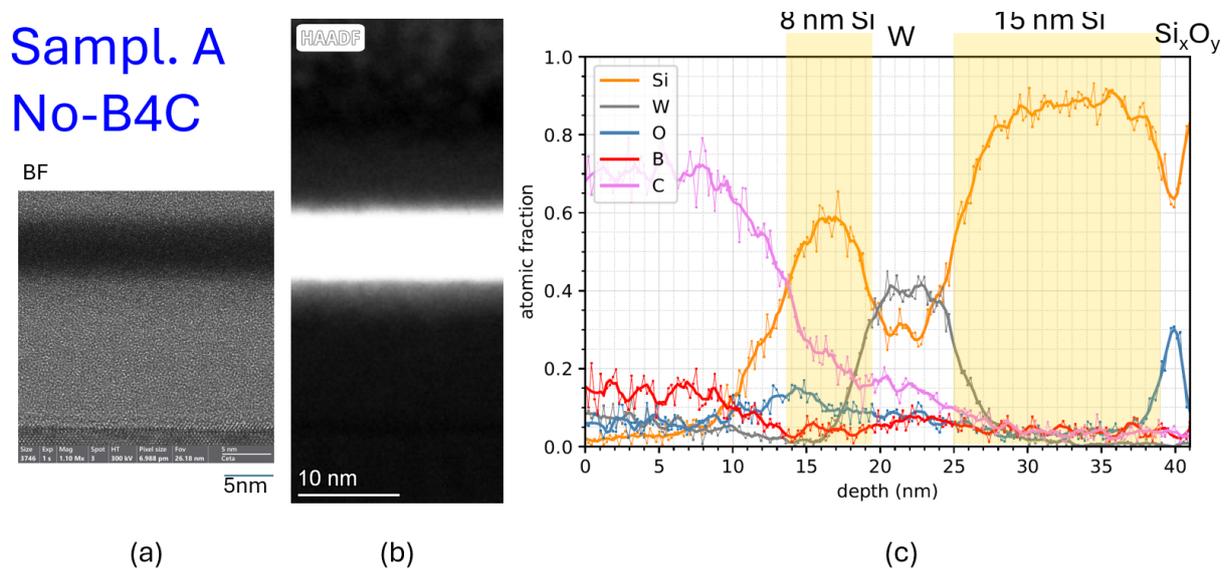

*Fig. S5. TEM analysis of sample A. In (a), a bright field (BF) image of the cross-section of the sample is displayed. The scale bar represents 5 nm. In (b), the high-angle annular dark-field image obtained by scanning-TEM is displayed. The scale bar is 10 nm. The plot in (c) displays the Energy Dispersive X-ray (EDX) analysis corresponding to the area in (b). The plot yields the atomic fraction of each element as a function of depth in the sample. The areas describing the deposited silicon films are highlighted in the plot as a yellow area.*

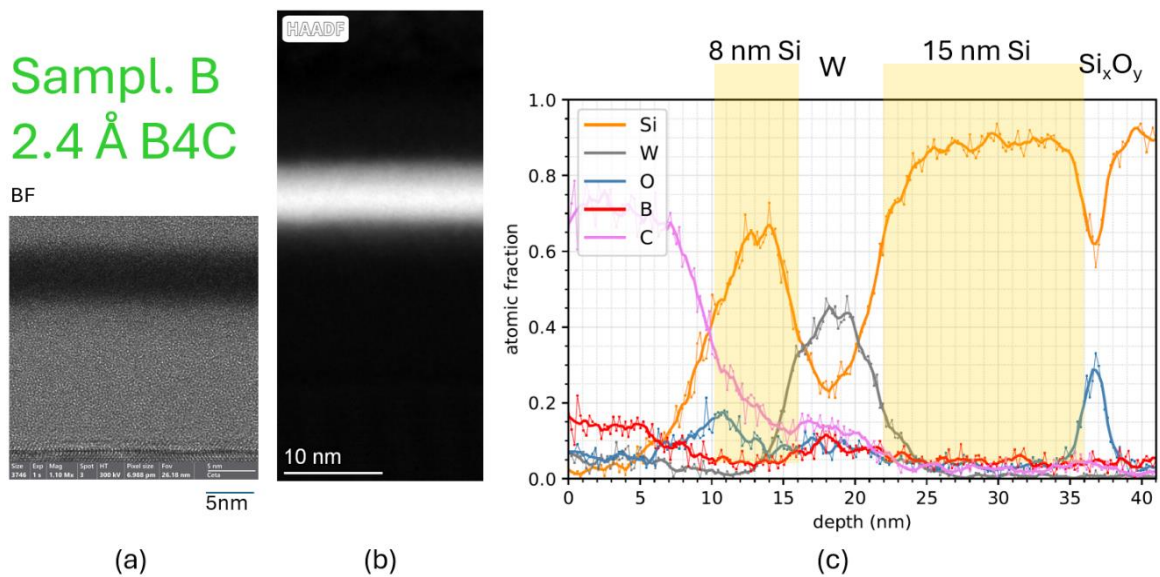

*Fig. S6. TEM analysis of sample B. In (a), a bright field (BF) image of the cross-section of the sample is displayed. The scale bar represents 5 nm. In (b), the high-angle annular dark-field image obtained by scanning-TEM is displayed. The scale bar is 10 nm. The plot in (c) displays the Energy Dispersive X-ray (EDX) analysis corresponding to the area in (b). The plot yields the atomic fraction of each element as a function of depth in the sample. The areas describing the deposited silicon films are highlighted in the plot as a yellow area.*